\documentclass[aps,prd,superscriptaddress,nofootinbib]{revtex4}
\usepackage{graphicx,epsfig,color}
\usepackage{bm}
\begin{document}
\title{Comment on ``Landau equation and QCD sum rules \\
for the tetraquark molecular states'' [Phys.~Rev.~{\bf D} 101, 074011 (2020)]}
\author{Wolfgang Lucha}
\affiliation{Institute for High Energy Physics, Austrian Academy of
Sciences, Nikolsdorfergasse 18,\\
A-1050 Vienna, Austria}
\author{Dmitri Melikhov}
\affiliation{D.~V.~Skobeltsyn Institute of Nuclear Physics,
M.~V.~Lomonosov Moscow State University,\\
119991, Moscow, Russia}
\affiliation{Joint Institute for Nuclear Research, 141980 Dubna, Russia}
\affiliation{Faculty of Physics, University of Vienna, Boltzmanngasse 5, A-1090 Vienna, Austria}
\author{Hagop Sazdjian}
\affiliation{Universit\'e Paris-Saclay, CNRS-IN2P3, IJCLab, 
91405 Orsay, France}

\date{\today}

\begin{abstract}
Recently, a QCD sum-rule analysis of tetraquark molecular states has
been published,
having the objective of demonstrating that our previously formulated
tetraquark-adequate QCD sum rules are not correct.
This comment brings to the attention of the reader inconsistencies in
that article and explains some subtle details
of QCD sum rules for exotic states.
\end{abstract}
\maketitle

\section{Sum rules for two-point functions of four-quark currents}
The basic point of Ref.~\cite{wang} is highlighted in the abstract:
{\it The tetraquark molecular states begin to receive contributions at the order $O(\alpha_s^0)$ rather than at the order $O(\alpha_s^2)$.}
Our recent papers \cite{lms1,lms2} have discussed the formulation of QCD sum rules for exotic states and have demonstrated that
$O(\alpha_s^0)$ and $O(\alpha_s^1)$ contributions on the OPE (Operator Product Expansion) 
side do not appear in the QCD sum rule 
appropriate for the tetraquark ($T$-adequate sum rules),
because of certain exact cancellations that occur between the OPE and the phenomenological sides of the sum rules for correlators of multiquark currents.
The results of \cite{wang} contradict our findings and this comment aims to
show that this is due to an inconsistency in the analysis of \cite{wang}.

In \cite{lms1,lms2} we have studied two-point functions of globally
color-singlet tetraquark currents consisting
of quark fields of four different flavors $a,b,c,d$. In the general case \cite{jaffe}, one can arrange the quark flavors
in these tetraquark currents in two different ways in terms of {\it color-singlet bilinears} of quark fields:
\begin{eqnarray}
\label{theta}
\theta_{\bar ab \bar c d}=(\bar q_a q_b)  (\bar q_c q_d) \mbox{ and }\theta_{\bar ad \bar cb}=(\bar q_a q_d)  (\bar q_c q_b).
\end{eqnarray}
All possible Dirac structures of these quark bilinears should be considered; this technical complication does not change anything
in the formulation of the $T$-adequate QCD sum rules and we do not write explicitly the Dirac structure between the quark fields.
By Fierz transformations, any other global four-quark color singlet can be expanded over the sum of currents of the 
two flavor structures in (\ref{theta}).
The tetraquark currents (\ref{theta}) lead to Green functions of two different topologies to be considered separately (direct and recombination channels) \cite{lms1,lms2}:
\begin{eqnarray}
\label{dir}
\Pi^{\rm dir}(x)=\langle {\rm T}\{\theta_{\bar ab\bar cd}(x) \theta^\dagger_{\bar ab \bar cd}(0)\}\rangle, \\
\label{rec}
\Pi^{\rm rec}(x)=\langle {\rm T}\{\theta_{\bar a b\bar cd}(x) \theta^\dagger_{\bar ad \bar cb}(0)\}\rangle.
\end{eqnarray}
Ref.~\cite{wang} considers only the direct Green function, Eq.~(\ref{dir}), that we have discussed in detail in \cite{lms1}.
To see the origin of the inconsistency in \cite{wang}, it is instructive to start with the four-point direct Green
function of the bilinear quark currents ($j_{\bar ij}=\bar q_i q_j$):
\begin{eqnarray}
\label{4pt}
\Gamma(x_1,x_2|y_1,y_2)=\langle T\{ j_{\bar ab}(x_1)j_{\bar cd}(x_2)  j^\dagger_{\bar ab}(y_1) j^\dagger_{\bar cd}(y_2)\}\rangle.
\end{eqnarray}
We can unambiguously split this quantity into factorizable and nonfactorizable parts
\begin{eqnarray}
\Gamma(x_1,x_2|y_1,y_2)=
\Gamma_{\rm fact}(x_1,x_2|y_1,y_2)+\Gamma_{\rm nonfact}(x_1,x_2|y_1,y_2),
\end{eqnarray}
where the factorizable part is defined as
\begin{eqnarray}
\Gamma_{\rm fact}(x_1,x_2|y_1,y_2)=
\langle T\{ j_{\bar ab}(x_1) j^\dagger_{\bar ab}(y_1)\} \rangle \langle T\{ j_{\bar cd}(x_2) j^\dagger_{\bar cd}(y_2)\} \rangle. 
\end{eqnarray}
In coordinate space, it is just the product of the two-point Green functions of the bilinear quark currents; in momentum space,
one has a convolution. Let us saturate the factorizable part by hadron intermediate states:
\begin{eqnarray}
\Pi_{\bar ab}(x)\equiv \langle T\{ j_{\bar ab}(x) j^\dagger_{\bar ab}(0)\} \rangle = \sum_{h_{\bar ab}}R_{\bar ab}(x),\\
\Pi_{\bar cd}(x)\equiv \langle T\{ j_{\bar cd}(x) j^\dagger_{\bar cd}(0)\} \rangle = \sum_{h_{\bar cd}}R_{\bar cd}(x),
\end{eqnarray}
where $R_{\bar ab}(x)$ and $R_{\bar cd}(x)$ are the quantities coming from
hadron saturation, the explicit form of which is irrelevant.
Important for us is the fact that the sum runs over the full system of hadron states 
with flavors $\bar ab$ ($h_{\bar ab}$) and ${\bar cd}$ ($h_{\bar cd}$), respectively. 
Consequently, the system of the intermediate hadron states that emerges in the factorizable
part of $\Gamma(x_1,x_2|y_1,y_2)$ is just the direct product of these two systems, $h_{\bar ab}\otimes h_{\bar cd}$.
The lowest representatives of $h_{\bar ab}$ and $h_{\bar cd}$ are one-meson states $M_{\bar ab}$ and $M_{\bar cd}$, respectively.
No other hadron state, in particular no exotic state, contributes here.\footnote{Ref.~\cite{wang} confuses
``non-interacting two-meson states'' and ``scattering two-meson states'':
the tetraquark state (molecular or compact) cannot be distinguished from ``scattering two-meson states'' on the
basis of the diagram analysis; but, of course, the tetraquark state can be easily distinguished from two non-interacting mesons.}  Consequently, we have to conclude that $\Gamma_{\rm fact}(x_1,x_2|y_1,y_2)$
has no relationship with exotic states. An exotic state, if it exists in the hadron spectrum of $\bar ab\bar cd$ states,
contributes only to $\Gamma_{\rm nonfact}(x_1,x_2,|y_1,y_2)$.

It is also clear that any other Green function, derived from $\Gamma_{\rm fact}(x_1,x_2,|y_1,y_2)$, knows nothing about the exotic state.
In particular, this is the case of the factorizable contribution to the
exotic two-point function, $\Pi^{\rm dir}_{\rm fact}(x)\equiv \Pi_{\bar ab}(x)\Pi_{\bar cd}(x)$, 
and the latter is therefore irrelevant for the QCD sum rule for
this state.
An appropriate QCD sum rule for the exotic state \cite{lms1} is based on the duality relation for $\Pi^{\rm dir}_{\rm nonfact}$, defined as 
$\Pi^{\rm dir}=\Pi^{\rm dir}_{\rm fact}+\Pi^{\rm dir}_{\rm nonfact}$.

It might be useful to emphasize at this place the qualitative difference between the correlator of the exotic four-quark currents
and the correlator of the bilinear quark currents.
Consider $\Pi_{\bar ab}(x)=\langle T\{j_{\bar ab}(x)j^\dagger_{\bar ab}(0)\}\rangle$.
Working in coordinate space, we may isolate the factorizable part of this function,
$\Pi_{\bar ab,{\rm fact}}=S_a(x)S^\dagger_b(x)$, where $S_i(x)$ is the (full) propagator of the quark with flavor $i$.
Now, we saturate $\Pi_{\bar ab}(x)$ with the full system of intermediate hadron states.\footnote{Here, QCD
confinement plays a crucial role: it ensures that free quarks do not appear as asymptotic states;
had free quarks and gluons existed, the duality relations would have been
looked at completely differently.}
Noteworthy, the intermediate one-meson state gives a nonzero contribution not only to the full $\Pi_{\bar ab}$,
but also to its factorizable part $\Pi_{\bar ab,{\rm fact}}$. [Here is the key difference with the exotic correlators:
in the latter case the tetraquark bound state does not appear in the set of the intermediate
states of the factorizable part of the correlation function.]
We then calculate $\Pi_{\bar ab}(x)$ using OPE, confront it against the hadron saturation, and
extract, \textit{via} technical procedures, the parameters of the ground state. Finally, the bulk of the
meson observables come from $O(\alpha_s^0)$ diagrams of $\Pi_{\bar ab,{\rm fact}}$.
This procedure is quite sensible, since the intermediate ground state gives a nonzero contribution
not only to the full $\Pi_{\bar ab}$, but also to $\Pi_{\bar ab,{\rm fact}}$.

In the correlator of the exotic currents, the factorizable part, $\Pi^{\rm dir}_{\rm fact}(x)$,
receives no contribution from the exotic intermediate state;
therefore, no information about this exotic state may be extracted from $\Pi^{\rm dir}_{\rm fact}$.
Since the perturbative $O(\alpha_s^0)$ and $O(\alpha_s)$
diagrams are exclusively of the factorizable type, they have no relationship with the exotic state.
The OPE side of the $T$-adequate QCD sum rule involves nonfactorizable contributions;
perturbative diagrams of this type start at order $O(\alpha_s^2)$ \cite{lms1}.

Having in mind these arguments, let us consider the derivation of the sum rule of \cite{wang}.
Constructing properly the OPE and the hadron sides,
one should observe an {\it exact cancellation} of the factorizable OPE contributions against the contributions of
non-interacting $h_{\bar ab}$ and $h_{\bar cd}$ hadron clusters on the phenomenological side. This property should follow by virtue of the QCD sum rules
for $\Pi_{\rm ab}$ and $\Pi_{\rm cd}$.
However, Ref.~\cite{wang} completely ignores the free two-meson
contribution on the phenomenological side, which is clearly distinct
from that of the two-meson scattering states, the latter contributing
to the nonfactoriazable part of the Green function (see also discussion in \cite{kondo}).
As a result, in the sum rule of \cite{wang}, the tetraquark receives
contributions from the $O(\alpha_s^0)$ perturbative diagrams, in
contradiction with the exact cancellation phenomenon of this term
with the free two-meson state contribution, as outlined above.

We now give a simple example showing that the sum rule of \cite{wang} is ``inadequate'' and leads also 
to a contradiction with some well-known results.
Let us apply the sum rule Eq.~(8) of \cite{wang} to a tetraquark in the $SU(N_c)$ gauge theory 
[QCD for a large number of colors $N_c$]. The $O(\alpha_s^0)$ diagram [Fig.~3 of \cite{wang}]
contains two color loops and thus scales as $N_c^2$. 
Following the procedure of \cite{wang}, one then obtains
\begin{eqnarray}
\label{fT}
f_T\equiv \langle T_{\bar ab \bar cd}|\theta_{\bar ab \bar cd}(0)|0\rangle\propto N_c.
\end{eqnarray}
As is well known \cite{witten,coleman}, one has for the ordinary mesons
\begin{eqnarray}
\label{fM}
f_M\equiv \langle M_{\bar ab}|j_{\bar ab}(0)|0\rangle\propto \sqrt{N_c}.
\end{eqnarray}
Combining (\ref{fT}) and (\ref{fM}), one would conclude that the four-quark current produces from the vacuum
the tetraquark state with the same strength as two-meson states. This is, however, in clear contradiction with the
well-known property that, at $N_c$-leading order, QCD Green functions are saturated by free meson states \cite{witten,coleman};
no exotic states appear at $N_c$-leading order.


\section{Conclusions}
\noindent
$\bullet$ There is a qualitative difference between the correlation
functions
of the ordinary bilinear or trilinear quark currents and the exotic multiquark (four-, five-, six-, etc.) currents.
This difference is related to the clusterization property of the multiquark operators \cite{lms3}.
For correlators of the ordinary currents, one may approximate the contribution of the hadron continuum by 
the {\it effective continuum} \cite{svz},
i.e., by the contribution of perturbative QCD (pQCD) diagrams above some
effective threshold. Consequently, the ground-state contribution is
related to
the low-energy region of pQCD diagrams, starting at $O(\alpha_s^0)$, complemented by condensate contributions.
In the low-energy region, the ground state gives the dominant contribution on the phenomenological side of the QCD sum rule;
this property is known in large-$N_c$ QCD and works well in the real world,
where $N_c=3$.

\vspace{.2cm}
In the correlation function of the exotic tetraquark currents, the situation is {\it qualitatively} different:
two-meson states always provide {\it the dominant contribution} on the phenomenological side of the QCD sum rule.
This property is rigorous in large-$N_c$ QCD, and holds also in real QCD.
Therefore, the Ansatz for the phenomenological side of a QCD sum rule in the form ``the tetraquark pole plus the effective continuum
equal to the pQCD diagrams above some effective threshold'', that worked
well for the ordinary correlation functions,
is inconsistent when applied to the exotic correlation functions.\footnote{At best, the standard Ansatz of a
``pole plus effective continuum'' may be applied to the nonfactorizable parts of the exotic correlation functions.}
One should first take into account the exact cancellations of factorizable diagrams on the OPE side against the contributions of
non-interacting hadron clusters $h_{\bar ab}$ and $h_{\bar cd}$ on the phenomenological side. When these cancellations
are not taken into account, one ends up with inconsistent QCD sum rules
\cite{wang}.

\vspace{.4cm}
\noindent
$\bullet$
We would also like to comment on a confusing statement of \cite{wang} concerning the Landau equations \cite{landau}:
{\it The Landau equation is of no use to study the Feynman diagrams in the QCD sum rules
for the tetraquark molecular states}.
Putting in question the use of the Landau equations in the context of QCD sum rules sounds strange:
Apart from condensate contributions, hadron observables in the method of QCD sum rules are related to low-energy integrals of the
imaginary parts of pQCD Feynman diagrams. As soon as one considers a Feynman diagram, Landau equations fully describe
the location of its singularities.

Concerning the relevance of the four-quark singularities of Feynman diagrams to the tetraquark, Ref.~\cite{wang} 
confused an important feature of the diagram selection formulated in \cite{lms1,lms2}: 
the criterion of selecting the tetraquark-relevant diagrams on the 
basis of their four-quark singularities applies to the four-point functions (\ref{4pt}), 
and not to the two-point function (\ref{dir}); 
the latter quantity, in addition to the tetraquark-relevant singularities, contains also four-quark singularites related 
to the non-interacting hadron clusters $h_{\bar ab}$ and $h_{\bar cd}$, not relevant to any tetraquark state. 

\acknowledgments{D.~M. and H.~S. gratefully acknowledge support by the joint RFBR/CNRS project 19-52-15022.
This work was done during the $stay@home$ international action to stop the COVID-19 epidemic.}


\end{document}